\documentclass[aps,preprint,nofootinbib,superscriptaddress,prc]{revtex4}
\usepackage{amssymb}

\usepackage{epsfig}
\usepackage[bookmarksnumbered,bookmarksopen,colorlinks,citecolor=blue,linkcolor=blue]{hyperref}
\begin{document}


\title{Driving Potential and Fission-Fragment Charge Distributions }

\author{Yuan Su }
\affiliation{Department of Physics, Guangxi Normal University,
Guilin 541004, P. R. China}

\author{Min Liu }
\affiliation{Department of Physics, Guangxi Normal University,
Guilin 541004, P. R. China}
\affiliation{Guangxi Key Laboratory of Nuclear Physics and Technology,
Guilin 541004, P. R. China}

\author{Ning Wang }
\thanks{wangning@gxnu.edu.cn}
\affiliation{Department of Physics, Guangxi Normal University,
Guilin 541004, P. R. China}
\affiliation{Guangxi Key Laboratory of Nuclear Physics and Technology,
Guilin 541004, P. R. China}

\begin{abstract}
We propose an efficient approach to describe the
fission-fragment charge yields for actinides based on the driving
potential of fissioning system. Considering the properties of
primary fission fragments at their ground states, the driving
potential, which represents the potential energies of the system around
scission configuration and closely relates to the yields of fragments,
can be unambiguously and quickly obtained from the Skyrme
energy-density functional together with the Weizs\"acker-Skyrme mass model. The
fission-fragment charge distributions for thermal-neutron-induced
fission and spontaneous fission of a series of actinides,
especially the odd-even staggering in charge distributions can be well
reproduced. Nuclear dynamical deformations and pairing corrections of fragments play an important role in the charge distributions.

\end{abstract}

\maketitle

\begin{center}
\textbf{I. INTRODUCTION}
\end{center}

Nuclear fission is a field of very intense studies for more than half
century \cite{Moll09,Qiao21,madland2006,Lammer2008,Schmidt2010,Rand,Liutj}.
One of the most interesting characteristics of neutron or heavy-ion
induced fission is the huge difference in the mass and charge
distribution of the fission fragments for different nuclei. The
investigation of the fission fragment yields is of great importance
not only for nuclear engineering, but also for understanding fission
process, testing nuclear models and exploring the role of fission
recycling as well as the structure of extremely neutron-rich nuclei
for the study of r-process in nuclear astrophysics
\cite{Arc,BH,Panov}. It is yet not very clear, until now,
how the original compound nucleus is transformed into a variety of
different fragments. It is qualitatively thought that the shell
effect plays a role in the double-humped distribution for the
fission of some actinides. Accurate predictions for the fission
fragment distribution especially the charge distribution of
actinides including the odd-even staggering \cite{Schmidt11} are
still urgently required.

To describe fission dynamics and fission barrier, some microscopic
or semi-empirical approaches, such as the Skyrme Hartree-Fock models
\cite{Bonn,Gor}, the covariant density functional theory
\cite{Lu,Kar} and the macroscopic-mircoscopic models
\cite{Moll09,Rand,Dob,Kow}, were established for calculating the
potential energy surface (PES) of a fissioning system from ground
state deformation to scission configurations. With the aid of modern
computer, the fission barriers of nuclei can be successfully
reproduced with a deviation of about one MeV \cite{Kar,Lu,Moll09},
and the fission fragment distributions can be roughly reproduced
based on the five-dimensional PES from the macroscopic-microscopic
model \cite{Rand}. The calculations of multi-dimensional PES are
time-consuming since millions of wave equations for strongly
(triaxial) deformed nuclear potentials need to be solved for
obtaining the single-particle levels of the system at different
deformation configurations. Furthermore, the determination of the
model parameters especially the strength of the spin-orbit
interaction and that of the pairing force from saddle to scission is
difficult in the traditional PES calculations, which could reduce the
predictive power of the models for describing the yields of
fission fragments. It is therefore necessary to develop an
alternative more efficient method for studying fission around
scission and post-scission movements.

To describe the competition between quasi-fission and complete fusion of super-heavy system, the di-nuclear system (DNS) concept was successfully proposed \cite{Adamian2003,Kalandarov2011,Nan,Nan12,Huang}. In fact the microscopic shell structure and even-odd effect can be involved in the potential energy surface of DNS. According to the DNS concept, each fission fragment around
the scission point retains its individuality in the evolution of the
DNS. In Ref. \cite{Sun}, it was found that the
valley of the driving potential for the mass number of heavy
fragments locates at $140$ for neutron-induced fission of $^{235}$U,
which is in good agreement with the peak of measured mass
distribution. It is therefore interesting to investigate the
charge yields for fission of actinides based on the
corresponding driving potentials.

In this work, we attempt to study the yields of fission fragments
based on DNS concept. Different from
the traditional studies based on the whole potential energy surface of a fissioning system from ground state to the scission point, we focus on potential energy surface around scission configuration. We would like to study the influence of nuclear structure effect on the yields of primary fission fragments especially the odd-even staggering in the charge
distribution for binary fission of actinides at low excitation
energies.

\begin{center}
\textbf{II. DRIVING POTENTIAL OF A FISSIONING SYSTEM}
\end{center}

The total potential energy of a fissioning system around scission is
written as
\begin{equation}
E_{\rm tot}=E_{1}(\vec{\beta} ) + E_{2}(\vec{\beta} ) + V(\vec{\beta},R),
\end{equation}
where, $E_{1}$ and $E_{2}$ denote the potential energies of the
light and heavy fission fragments respectively, which are functions of nuclear deformations. $V(\vec{\beta},R)$ denotes the
interaction potential between two fragments with the center-to-center distance $R$, which is obtained
with the Skyrme energy-density functional plus the extended Thomas-Fermi (ETF) approximation
\cite{Liu}. In the Skyrme energy-density functional approach, the total binding energy of a nucleus can be expressed as the integral of energy density functional which is a function of nuclear densities of protons and neutrons under the ETF2 approximation \cite{Bart82}.
In our calculations, the nuclear central densities and surface diffuseness for a certain nucleus (or fragment) are firstly determined by using the restricted density variational method and taking the neutron and proton density distributions as spherical symmetric Fermi functions \cite{Liu}. Then, we introduce nuclear deformations $\vec{\beta}$ in the radius parameter, remaining the central densities and surface diffuseness of the nucleus (or fragment) unchanged, to consider the influence of nuclear deformations on the interaction potential between two fragments.

Assuming that a compound nucleus separates into a certain pair of fission fragments, $ (A_{\rm CN},Z_{\rm
CN})\rightarrow (A_1,Z_1) + (A_2,Z_2)$, with mass number $A_{\rm CN}=A_1+A_2$
and charge number $Z_{\rm CN}=Z_1+Z_2$ in the fission process, we define the
driving potential of the fissioning system as,
\begin{equation}
U(A_{\rm f},Z_{\rm f}) = E_{\rm tot} - E_{\rm CN} = Q_{g.s.}+\Delta
Q ( \vec{\beta} ) + V(\vec{\beta},R )
\end{equation}
with mass number $A_{\rm f}$ and charge number $Z_{\rm f}$ for one
fragment. The driving potential describes the potential energy of
the fissioning system around scission configuration. $E_{\rm CN}$ and $Q_{g.s.}$
denote the energy of the compound nucleus at its ground state
and the $Q$ value of the reaction system,
respectively. $\vec{\beta}$ denotes the deformations of fission
fragments around scission configuration. In this work, the
static deformations for each
fission fragment, i.e., the deformations for nuclei
at their ground states, are determined by Weizs\"acker-Skyrme (WS) mass model \cite{WS33}. Simultaneously, the dynamical
deformations $\beta^D$ of fragments are also considered. $\Delta
Q=[E_1(\beta^D)-E_1^{g.s.}]+[E_2(\beta^D)-E_2^{g.s.}]$ denotes the
change of the potential energies for the fission fragments with
respect to the individual energies at their ground states due
to the dynamical deformations.

For description of the elongated tip-tip structure of a fissioning system around scission point, nuclear prolate shapes and the octupole deformations should play a dominant role. In this work, we take the absolute value $|\beta_2|$ for nuclei with oblate deformations and set the values of dynamical octupole deformation $\beta_3^D$ for a pair of fission fragments to be the same but in reverse for simplicity. By
varying the value of $\beta_3^{D}$ for a certain pair of fission
fragments at the position of potential barrier where the two fragments are slightly separated and searching for the minimal value of the driving potential $U$, one can obtain the optimal value for $\beta_3^D$
which is about $0.1\sim 0.2$ (one can find the similar result from Fig.10 in Ref.\cite{Wang07}).
For fragments with spherical shapes at their ground states, the dynamical quadrupole deformations
$\beta_2^{D}$ are also considered, and the values of $\beta_2^{D}$
are obtained like those for $\beta_3^{D}$.

From Eq.(2), one can see that in addition to the interaction potential, both the properties of fragments at their ground state and the dynamic deformations around scission configuration influence the potential energy surface of a fissioning system. The microscopic shell effect and pairing effect are effectively involved in the calculations of $Q_{g.s.}$ and the residual part $\Delta Q$. The values of $Q_{g.s.}$ are mainly taken from the measured masses with high accuracy in AME2016 \cite{AME2016a,AME2016b}. For the masses of unmeasured nuclei and the value of $\Delta Q$, we use the predictions of the WS mass models \cite{WS33,WS4}. We would like to emphasize that the driving potential $U$ for a certain fissioning system can be unambiguously obtained from the Skyrme energy-density functional and the WS
mass models, without introducing any new model parameters.

\begin{center}
\textbf{III. RESULTS AND DISCUSSIONS}
\end{center}

 \begin{figure}
\includegraphics[angle=-0,width= 0.8\textwidth]{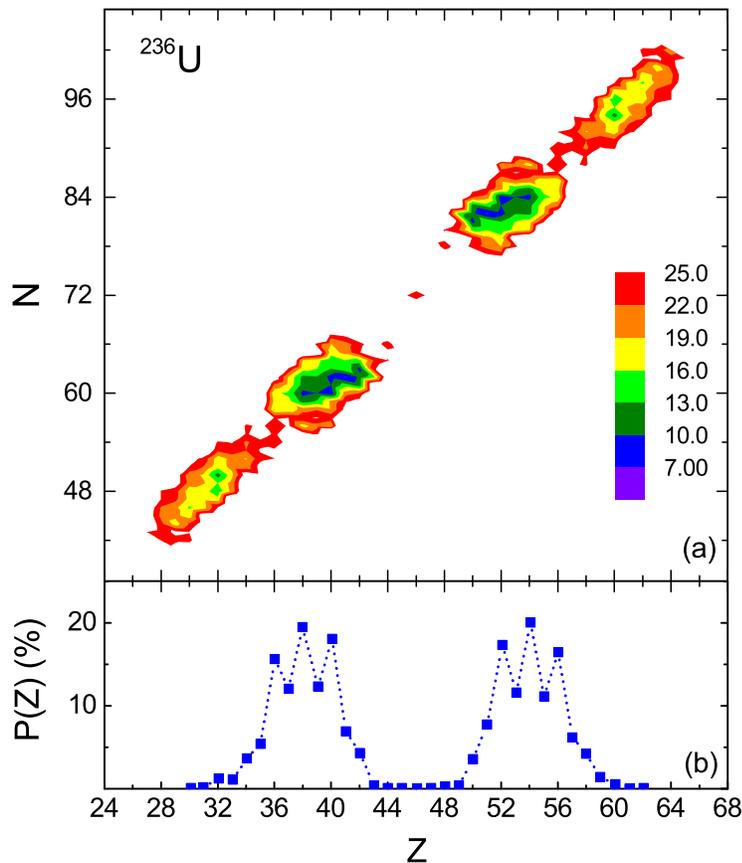}
 \caption{(Color online) (a) Calculated driving potential (in MeV) for fission of
$^{236}$U. (b) Charge yields of fission fragments for $^{235}$U(n$
_{\rm th}$, f) \cite{Chad}. }
\end{figure}

Considering that the mass and charge number of fission fragments are distributed at a wide range, the driving potential $U(A_{\rm f},Z_{\rm f})$ of a certain compound nucleus separating into different pairs of fission fragments is investigated. For example, the driving potential for fission of $^{236}$U with about five hundreds different pairs of fission fragments is calculated and the results are shown in Fig.1(a). Fig.1(b)
shows the data for the charge distribution of primary fission fragments
in thermal-neutron-induced fission of $^{235}$U(n$_{\rm th}$, f) \cite{Chad}.
One sees from the driving potential that there exists
two deep valleys located around $(A_{\rm f}=140,Z_{\rm f}=54)$ and
$(A_{\rm f}=96,Z_{\rm f}=38)$ which exactly respond to the peaks of
the mass and charge distributions of fission fragments. It indicates that
the yields of fission fragments of $^{236}$U at low excitation
energies are closely related with the corresponding driving
potential.

\begin{figure}
\includegraphics[angle=-0,width= 1\textwidth]{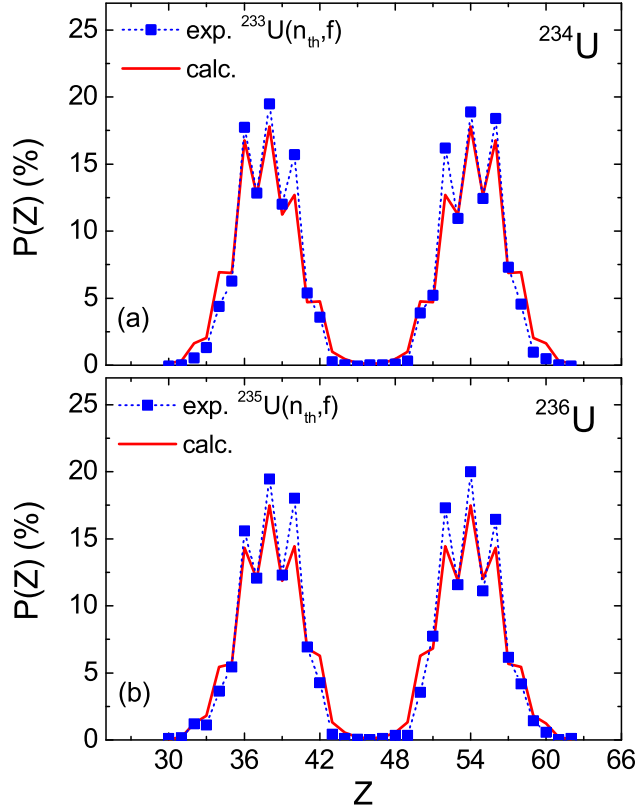}
 \caption{(Color online) Calculated and measured charge yields for fission of
$^{234}$U and $^{236}$U. The squares denote the experimental data for thermal-neutron-induced
fission of $^{233}{\rm U(n_{\rm th},f)}$ and $^{235}{\rm U(n_{\rm th},f)}$ \cite{Chad}. The red curves denote the calculated results with Eq.(3).}
\end{figure}

Based on the calculated driving potential, the corresponding fragment charge distributions for a fissioning system are further predicted. In this work, the yields of fission fragments for a fissioning
system is expressed as,
\begin{equation}
Y(A_{\rm f},Z_{\rm f})=C\exp \left [-\frac{U(A_{\rm
f},Z_{\rm f})}{K} \right ].
\end{equation}
$K$ is a model parameter, which relates to the temperature of the system around scission point. Here, we empirically set $K=0.38A_{\rm CN}-83.35$ (in MeV) for describing the fragment yields in fission of actinides at low
excitation energies. The corresponding coefficients in $K$ are determined by the measured fission-fragment charge distributions for a series of actinides. The normalization factor $C(A_{\rm f})=P(A_{\rm f})/\sum\limits_{Z_{\rm f}} \exp[-U/K]$ can be uniquely determined by a given fragment mass distribution $P(A_{\rm f})$. In the present calculations, we use the measured fragment mass distribution for the determination of the normalization factor $C$. If $P(A_{\rm f})$ is not available, one could use the empirical fission potential \cite{Sun14} for calculating the mass distributions. With the driving potential and mass distribution, the charge distributions of the primary fission fragments $P(Z_{\rm f})= \sum\limits_{A_{\rm f}}  {Y(A_{\rm f},Z_{\rm f})}$ can be directly calculated with Eq.(3). Here, we would like to emphasize that Eq.(3) is only applicable for describing the thermal-induced and spontaneous fission of actinides. If one would like to extend this model to describe the distributions of fission fragments and yields at high excitation energies, the temperature dependence for the driving potential $U$, the normalization factor $C$ and the model parameter $K$ should be considered.

Fig. 2 shows the calculated charge distribution of primary fragments for fission of
$^{234}$U and $^{236}$U. The experimental data for thermal-neutron-induced fission of $^{233}{\rm U(n_{\rm th},f)}$ and $^{235}{\rm U(n_{\rm th},f)}$, especially the odd-even staggering can be well reproduced.  From Eq.(2), one can see that nuclear deformations influence the value of $\Delta Q$ and the interaction potential $V$ in the calculations of the driving potential. To see the influence of nuclear deformations on the charge distributions, we compare the calculated results with nuclear ground state deformations from the WS model \cite{WS33} and those from the finite range droplet model (FRDM) \cite{Moll95}. In Fig. 3, we show the calculated charge distribution of primary fragments for thermal-neutron-induced fission of $^{234}$U, $^{236}$U, $^{240}$Pu and spontaneous fission of $^{252}$Cf. The red curves and the green cirlces denote the calculated results from the WS model and the FRDM model, respectively. Here, the data for the mass distributions of $^{233}$U(n$_{\rm th}$, f) \cite{U233}, $^{235}$U(n$_{\rm th}$, f) \cite{U235}, $^{239}$Pu(n$_{\rm th}$, f) \cite{Pu239} and spontaneous fission of
$^{252}$Cf \cite{Cf252mass} are adopted for the determination of the normalization factors $C$, respectively. One sees that with nuclear deformations from both mass models, the fragment charge distributions for fission of $^{234}$U, $^{236}$U, $^{240}$Pu and $^{252}$Cf can be reasonably well reproduced. The difference due to the predicted nuclear deformations from the two models can also be obviously observed for light fragments with $Z=42$. The yields for fragments with $Z=42$ and the corresponding parters based on the deformations from the FRDM model is much higher than those from the WS model, especially for fission of U and Pu. According to the driving potential shown in Fig. 1, the most probable neutron number is around $N=60$ for fragments with $Z=42$. We note that the predicted quadrupole deformation ($\beta_2=0.329$) from the FRDM model is much larger than that from the WS model ($\beta_2=0.210$) for $^{102}$Mo. The quadrupole deformation of nuclei can significantly affect the interaction potential $V(\vec{\beta},R)$ between fragments. With larger quadrupole deformation, one obtains lower potential barrier and thus lower driving potential which results in higher yields in the charge distributions.

\begin{figure}
\includegraphics[angle=-0,width= 1\textwidth]{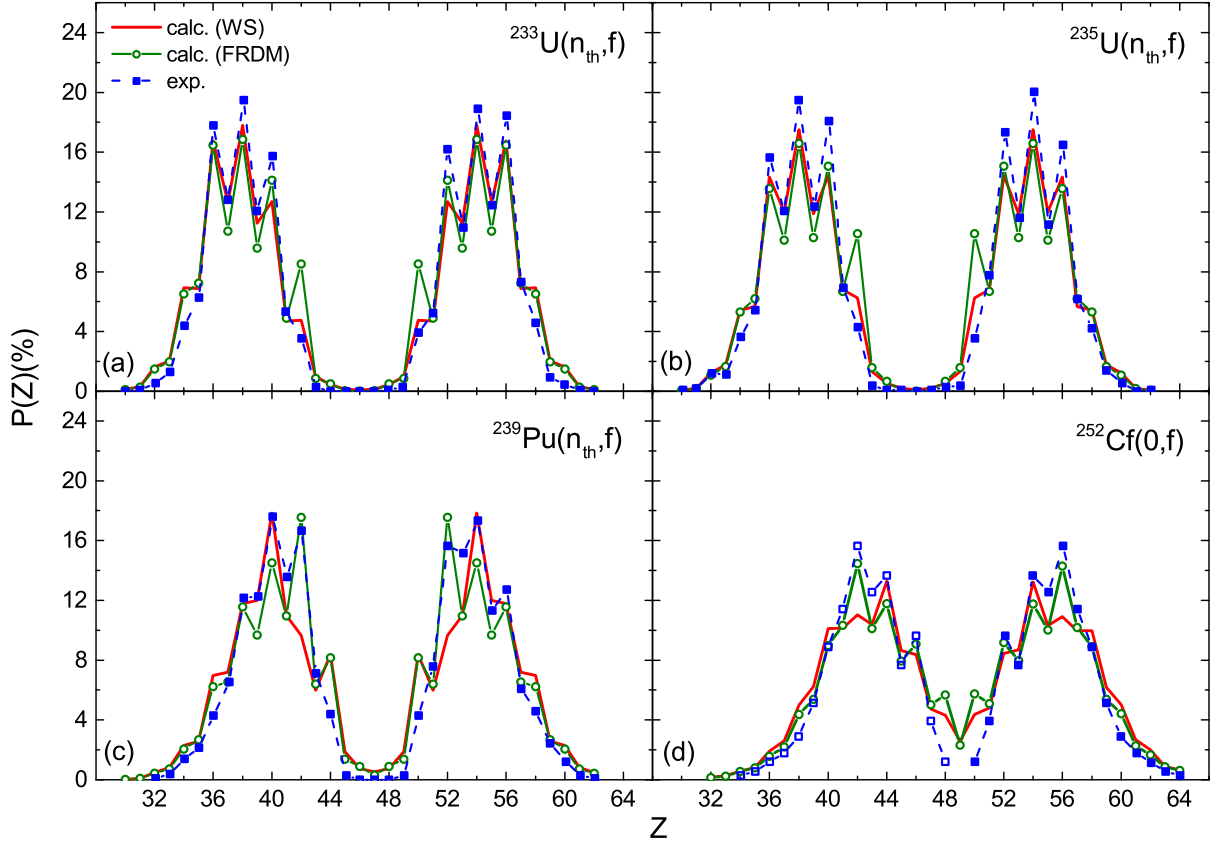}
 \caption{(Color online) Calculated and measured charge yields for thermal-neutron-induced fission of
$^{234}$U, $^{236}$U, $^{240}$Pu and spontaneous fission of $^{252}$Cf. The squares denote the experimental data taken from \cite{Chad,Cf252b}. The red curves and the green cirlces denote the calculated results with nuclear ground state deformations from the WS model and the FRDM model, respectively. The open squares in (d) denote the charge distributions obtained from the mirror of the measured heavy fragments.}
\end{figure}

In Ref. \cite{Rand}, Randrup and M\"oller calculated the charge yields for fission of $^{234,236}$U and $^{240}$Pu based on
Brownian motion on the five-dimensional potential-energy surfaces (5D-PES). We note that the odd-even staggering in the experimental data is not reproduced at all with their approach. To understand the physics behind the odd-even staggering in charge distribution, we explore the influence of pairing effect on the charge distribution. In the calculations of $Q_{g.s.}$ in Eq.(2), we remove the contribution of the pairing term which is expressed as $a_{\rm {pair}} A^{-1/3}\delta_{np}$ in the WS model \cite{WS33}, with
\begin{eqnarray}
\delta_{np}= \left\{
\begin{array} {r@{\quad:\quad}l}
  2 - |I|  &   N {\rm ~and~} Z {\rm ~even }    \\
      |I|  &   N {\rm ~and~} Z {\rm ~ odd }    \\
  1 - |I|  &   N {\rm ~even,~} Z {\rm ~odd,~ } {\rm ~and~ } N>Z   \\
  1 - |I|  &   N {\rm ~odd,~} Z {\rm ~even,~ } {\rm ~and~ } N<Z   \\
  1             &   N {\rm ~even,~} Z {\rm ~odd,~ } {\rm ~and~ } N<Z   \\
  1             &   N {\rm ~odd,~} Z {\rm ~even,~ } {\rm ~and~ } N>Z   \\
\end{array} \right.
\end{eqnarray}
The corresponding calculation results for fission of $^{236}$U are shown in Fig. 4. One sees that the odd-even staggering disappears, which clearly indicates that the pairing effect in fragments at their ground state plays a key role to the odd-even staggering in the charge distribution.

\begin{figure}
\includegraphics[angle=-0,width= 0.7\textwidth]{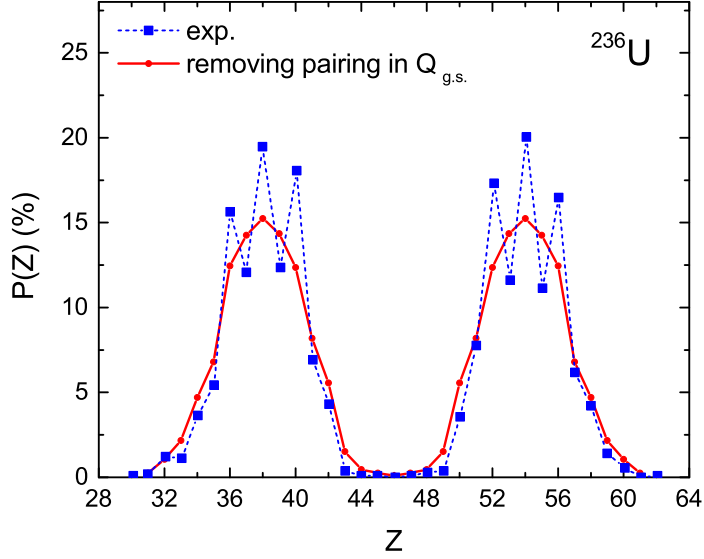}
 \caption{(Color online) The same as Fig. 2(b) but removing the pairing term in the calculations of $Q_{g.s.}$.}
\end{figure}

To understand the physics behind the model parameter $K$ in Eq.(3), we investigate the change of the $Q$ values for eight actinides, $^{230}$Th, $^{234,236}$U, $^{240,242}$Pu, $^{246}$Cm and $^{250,252}$Cf, from asymmetric fission to the corresponding symmetric ones. Considering that one peak of the fission-fragment mass distribution is usually located at about $A_{\rm f}=140$ and the corresponding peak of charge distribution is about $Z_{\rm f}=54$ for these actinides at low excitation energies, we calculate the corresponding ground-state $Q$ value, defined as $Q_{g.s.}^{\rm asym}$, with one fragment being $^{140}$Xe. In Fig. 5, we show the the difference between the $Q$ value for the asymmetric fission and that for the symmetric one $Q_{g.s.}^{\rm sym}$. One can see that the increasing trend of $Q_{g.s.}^{\rm asym}-Q_{g.s.}^{\rm sym}$ with mass number is very close to that of $K$. One could note that there exists a shift of 4.4 MeV to $Q_{g.s.}^{\rm asym}-Q_{g.s.}^{\rm sym}$ in Fig. 5, which relates to the degree-of-freedoms, e.g. the diffuseness and deformations of fission fragments, adopted in the calculations of the driving potential. We also note that if taking $K=Q_{g.s.}^{\rm asym}-Q_{g.s.}^{\rm sym}+4.4$ in the calculations, the obtained fragment charge distributions are comparable to those with $K=0.38A_{\rm CN}-83.35$. It seems that the value of $K$ may have a relationship with the difference of the $Q$ value.

\begin{figure}
\includegraphics[angle=-0,width= 0.7\textwidth]{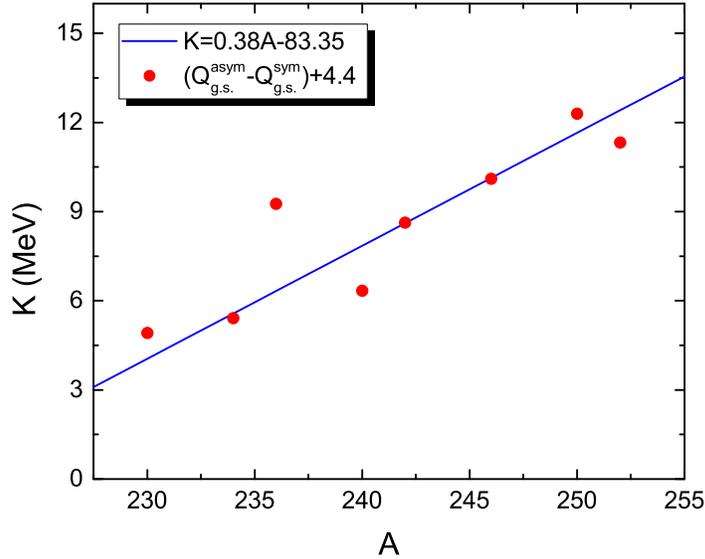}
 \caption{(Color online) Comparison of the model parameter $K$ and the change of $Q$ value from asymmetric fission to the symmetric one.}
\end{figure}

\begin{center}
\textbf{IV. SUMMARY}
\end{center}

In summary, the description of fission-fragment charge distribution of actinides can be
significantly improved with the driving potential of fissioning
system which creates a bridge between the mass distribution
and the charge distribution. Since a number of fission-fragment charge yields have not yet been measured while the corresponding mass distributions are available, the predictions for charge distributions based on the driving potential are quite useful and interesting. Considering the properties of primary
fission fragments at their ground states, the potential energies of
systems around scission configuration can be unambiguously and
quickly obtained from the Skyrme energy-density functional and
the WS mass models. The odd-even effect in the charge
distributions which links nucleon transfer through the neck in the
regime of strong pairing correlations is much better reproduced,
comparing with the traditional potential energy-surface approach in
which the determination of the model parameters especially the
strength of pairing force becomes difficult for the extremely
deformed shapes. There are two advantages in the proposed approach:
1) The CPU time is significantly shortened because only thousands (in
the calculations of $\Delta Q$) rather than millions of wave equations
need to be solved to obtain the potential energies of the system in
the calculation of the fission-fragment yields. 2) The microscopic shell and pairing
effects are more accurately taken into account via the measured $Q$ value and the residual term $\Delta Q$ which are calculated by using the WS mass models with high accuracy for describing the known masses \cite{Li21}.
We find that the structure effect of fragments at their ground state, such as the deformations and the odd-even staggering play a crucial role for the fission of actinides around
scission at low excitation energies.

\begin{center}
\textbf{ACKNOWLEDGEMENTS}
\end{center}

This work was supported by National Natural Science Foundation of China (Nos
U1867212, 11875323, 12147211) and Guangxi Natural Science Foundation (No. 2017GXNSFGA198001).

\end{document}